# Financial Conduct Authority
*Occasional Paper 43*  *July 2018*

# Weighing anchor on credit card debt

*Benedict Guttman-Kenney, Jesse Leary and Neil Stewart*

# FCA occasional papers in financial regulation

**The FCA occasional papers**

The FCA is committed to encouraging debate on all aspects of financial regulation and to creating rigorous evidence to support its decision-making. To facilitate this, we publish a series of Occasional Papers, extending across economics and other disciplines.

The main factor in accepting papers is that they should make substantial contributions to knowledge and understanding of financial regulation. If you want to contribute to this series or comment on these papers, please contact Karen Croxson at karen.croxson@fca.org.uk.

**Disclaimer**

Occasional Papers contribute to the work of the FCA by providing rigorous research results and stimulating debate. While they may not necessarily represent the position of the FCA, they are one source of evidence that the FCA may use while discharging its functions and to inform its views. The FCA endeavours to ensure that research outputs are correct, through checks that include independent referee reports, but the nature of such research and choice of research methods is a matter for the authors using their expert judgement. To the extent that Occasional Papers contain any errors or omissions, they should be attributed to the individual authors, rather than to the FCA.

**Authors**

Benedict Guttman-Kenney and Jesse Leary work at the Financial Conduct Authority.

Neil Stewart is a Professor of Behavioural Science at Warwick Business School.

**Acknowledgements**

We are grateful to the institution we worked with for their cooperation and patience, without which this research would not have been possible. We would also like to thank the many FCA staff who have contributed at various stages to this research – especially Mary Starks and Stefan Hunt for their support. We are extremely appreciative of the feedback from Paul Adams and Paolina Medina.

This research was supported by grants from the Economic and Social Research Council ES/K002201/1, ES/P008976/1, ES/N018192/1, and the Leverhulme Trust RP2012-V-022, awarded to Neil Stewart.  These funding sources provided financial support but were not involved in any other aspects of the research.

All our publications are available to download from www.fca.org.uk. If you would like to receive this paper in an alternative format, please call 020 7066 9644 or email publications_graphics @fca.org.uk or write to Editorial and Digital Department, Financial Conduct Authority, 12 Endeavour Square, London E20 1JN.

# Contents

# Abstract

We find it is common for consumers who are not in financial distress to make credit card payments at or close to the minimum. This pattern is difficult to reconcile with economic factors but can be explained by minimum payment information presented to consumers acting as an anchor that weighs payments down. Building on Stewart (2009), we conduct a hypothetical credit card payment experiment to test an intervention to de-anchor payment choices. This intervention effectively stops consumers selecting payments at the contractual minimum. It also increases their average payments, as well as shifting the distribution of payments. By de-anchoring choices from the minimum, consumers increasingly choose the full payment amount – which potentially seems to act as a target payment for consumers. We innovate by linking the experimental responses to survey responses on financial distress and to actual credit card payment behaviours. We find that the intervention largely increases payments made by less financially-distressed consumers. We are also able to evaluate the potential external validity of our experiment and find that hypothetical responses are closely related to consumers' actual credit card payments.

# 1   Introduction

One in four UK credit card payments are at, or very near to, the contractual minimum (FCA, 2016). Why? Are the finances of one in four credit card users so constrained that they can only afford to pay the minimum amount due? Or is there some other factor causing them to choose it? We investigate this using a large, bespoke survey of credit card users with a hypothetical payment experiment matched to their actual credit card payments. Different explanations for payment choices have very different implications for interventions to increase credit card payments. If consumers are financially constrained then it will be hard to increase their payments, but if they are not constrained then attempts to increase payments may be more fruitful.

The minimum payment on credit cards is prominently displayed as an easy-to-select option when consumers are making decisions on how much credit card debt to repay. In a lab experiment, Stewart (2009) found that removing minimum payment information increases the amount consumers made in credit card payments. Stewart argued that minimum payment information act as a psychological anchor that reduces payments. Anchoring happens when the presence of irrelevant information biases people's decisions or judgements (Ariely, Lowenstein, & Prelec, 2003; Mussweiler, & Strack, 2000; Kahneman, 1973; Tversky, & Kahneman, 1974; Wilson, Houseton, Etling, & Brekke, 1996) and does not vanish with higher cognitive ability (Bergman, Elingsen, Johannesson, & Svensson, 2010). Stewart's lab results showed that minimum payment information not only appears to make people more likely to pay exactly the minimum but also weighs down payments close to it and, surprisingly, makes people less likely to pay the debt in full. These findings have since been replicated across a variety of studies with consumers in different countries over time (Adams, Guttman-Kenney, Hayes, Hunt, & Stewart, 2018; Jiang & Dunn, 2013; Navarro-Martinez et al., 2011).

The '*anchoring effects*' of minimum payments have been found beyond the lab. Previous studies have used data from the UK (Navarro-Martinez et al., 2011), US (Agarwal, Chomsisengphet, Mahoney, & Stroebel, 2015; Keys & Wang, 2017) and Mexico (Medina & Negrin, 2017) to measure the effects of changes in minimum payments on the full distribution of payments. These show effects not just on consumers forced by a higher minimum to make a higher payment, but also on the distribution of payments above the minimum. This indicates the minimum payment is acting as a psychological anchor. Information showing how long it will take to repay debt under alternative scenarios appears insufficiently powerful to combat such an anchoring effects (Adams, Guttman-Kenney, Hunt, Hayes, Laibson, & Stewart, 2018; Navarro-Martinez et al., 2011). It also possibly even produces small perverse effects, with these scenarios acting as targets that consumers aim. While such, seemingly helpful, information increases the payments some consumers make, it reduces the amounts paid by others (Agarwal et al., 2015; Hershfield & Roese, 2015; Keys & Wang, 2016; Salisbury, 2014; Soll, Keeney, & Larrick, 2013).

Recent work by Bartels and Sussman (2016) argues that the minimum payment acts as a target, rather than an anchor. With targets, consumers motivate themselves to reach the target value (Allen, Dechow, Pope, & Wu, 2016; Heath, Larrick, & Wu, 1999; Kahneman, 1992), unlike anchors where there is not an emotional response. By comparing credit card payment distributions to those in other contexts considered to be anchors (eg how estimates of the length of Mississippi river vary based on framing whether it is shorter or longer than particular distances) or targets (eg marathon finish times relative to goals), they conclude that minimum payments act more like a target than an anchor. Irrespective of whether minimum payments act as anchors or targets, both behavioural mechanisms result in the same outcomes when minimum payment information is displayed: making it more likely that consumers will pay the minimum or a value close to it.

So minimum payment information could be considered a bad nudge. Indeed, Thaler and Sunstein (2008) write in their influential book *Nudge* 'Credit cards minimum payment…can serve as anchor and as a nudge that this payment is an appropriate amount'. There is evidence to support this: a consumer survey of UK credit card holders who reported they made minimum payments 48% said that they thought minimum payments were an amount recommended by their credit card provider and a similar amount (50%) thought it was the amount most people chose to pay (Which?, 2015).

But are economic forces actually driving the clustering of payments around the contractual minimum? One hypothesis is that all these consumers are liquidity constrained (Agarwal, Liu & Souleles, 2007; Agarwal et al., 2015; Agarwal, Chomsisengphet, Mahoney, & Stroebel, 2017; Gross & Souleles, 2002; Leth-Petersen, 2010) and cannot afford to pay more than the minimum.[1] If this is the case, these consumers may want to pay more but be unable to do so. In section 2 we provide new, descriptive evidence that consumers paying at or close to the contractual minimum are not financially distressed. Instead they appear to being anchored to the minimum payment. This finding provides reason to test an intervention to de-anchor credit card payments.

We trial an intervention similar to that used in Stewart (2009) to de-anchor payment choices from the minimum. We conducted this experiment as part of an online survey of credit card consumers (we discuss details on the survey, including the sampling frame and response rate, in section 2). We present consumers with a hypothetical credit card bill and vary whether the minimum payment is included. This affects the extent to which consumers must make an active choice of their credit card payment amount by varying the way options are displayed – the '*choice architecture*' (Johnson et al., 2012; Keller, Harlam, Loewenstein & Volpp, 2011; Thaler, Sunstein, & Balz, 2014). We gave the control group a hypothetical online credit card payment scenario. This included the payment option radio buttons (full payment, minimum payment, free text box for other payment amount) and disclosure of information (statement balance due, minimum payment amount) they would normally see in the real-world. The treatment group were presented with a similar scenario, but with the minimum payment radio button option and the minimum payment amount removed. In either case, if a consumer entered an amount less than the contractual minimum, a prompt appeared that showed the contractual minimum and asked the consumer to re-enter their payment amount.

---

[1] Another hypothesis may be consumers are repeatedly only making minimum payments due to inertia of having automatic minimum payments set up which do not require the consumer to actively engage with their credit card (Sakaguchi, Stewart & Gathergood, 2018). Such behaviour is explored in Adams, Guttman-Kenney, Hayes, Hunt, Laibson & Stewart (2018) and Adams, Guttman-Kenney, Hayes, Hunt, Laibson & Stewart (2018) which respectively test the ability of disclosure and choice architecture to increase credit card payments for people who would otherwise be on automatic minimum payments.

This experiment builds on Stewart's (2009) experiment and adds to the existing literature in three ways. First, we have a much larger and more representative sample than earlier studies, which primarily relied on online panels. Our sample is of UK consumers who have recently taken out a credit card, and includes consumers with a wide range of credit risks and limits. Second, we test consumer payment choices in a low and a high balance scenario. Varying balance is important given the broad range of credit card debts - the anchoring effects of a £10 minimum payment on a £500 balance may differ from that of an £80 minimum payment on a £3,000 balance.

Our third contribution comes from our ability to link experimental and survey responses to administrative data on respondents' actual credit card usage.[2] This enables us to evaluate how likely responses to our hypothetical experiment are to translate to real-world behaviour. It also enables us to examine how average treatment effects vary based on consumers' actual credit card payment behaviour and circumstances.

We discuss our main experimental findings in section 3. In line with previous literature, we find that minimum payment information (amounts and options to pay the minimum) have large effects on consumer payment choices. We find that removing the explicit option to pay the minimum and the minimum payment amount has large effects on the distribution of payments. It reduces minimum payments and increases full payments and average payments. We find that these effects are directionally consistent across both low and high balance scenarios. By de-anchoring choices from the minimum, the full payment now seems to be chosen noticeably more often – possibly acting as a target payment amount for consumers to aim to pay.

In section 4, we develop approaches for evaluating the external validity of our survey experiment. We find that hypothetical responses are closely related to the same consumers' actual credit card payments. This gives us reassurance that if our intervention was applied to the real-world then the impacts may be similar to those found in our hypothetical experiment.

Finally, in section 5, we examine how effects vary across consumers. We find some evidence of effects varying based on the degree of self-reported financial distress. Broadly the less financially distressed the consumer, the larger the effect of the intervention. However, we do not have sufficient statistical power to say this conclusively as the number of consumers reporting to be most financially distressed is quite small and therefore the confidence intervals of our estimates for this group are wide. Financial constraints vary over time, so we would expect this intervention to result in consumers paying more in 'good times', which would result in less credit card debt and therefore lower contractual minimums payments in 'bad times' when they have fewer available funds. We can conclusively say that there appears to be strongly significant effects of the intervention increasing payments for those not reporting to be in financial distress.

---

[2] As discussed below, this information was only linked when participants gave explicit permission. Neither the FCA nor the researchers had access to any direct personal identifiers of the participants.

# 2   Survey design & motivation

## Survey design

In cooperation with a large UK-based credit card issuer, we conducted a survey of consumers who had recently taken out new credit cards. The target sample was consumers who had taken out a new credit card with this issuer between January and May 2017. The survey was sent by email, with the group encouraged to participate through a prize draw.[3] We ran a small pilot to refine the survey format and do not use those responses for analysis. The main survey took place in May 2018 and achieved 8,490 responses –a 4.3% response rate. The card issuer gave us data on whether consumers had automatic 'Direct Debits' payments in place, and, if so, whether it was set to pay the minimum, the full balance, or a fixed amount. The issuer also categorized customers into ten categories by average balance, from which we could not identify individual respondents.

We included a question in the survey asking respondents for consent to anonymously match individual survey responses to their credit card and credit file data. We observe data on each of the first seven statement cycles (including balances, spending and payments) since taking out a new card. 76.1% of respondents – 6,462 consumers – gave consent for us to match these data. These data were gathered by the UK financial regulator — the Financial Conduct Authority (FCA).

We begin our analysis taking all respondents who gave consent to match, then apply two restrictions based on their latest statement of administrative data observed.

   i.   First we remove all respondents who had an automatic payment set up (known as 'Direct Debits' or 'autopay') – reducing the sample to 3,285. We apply this restriction because consumers on automatic payments make payments in quite different ways to other consumers – they do not need to consciously make a payment each month. Two other papers (Adams, Guttman-Kenney, Hayes, Hunt, Laibson, & Stewart, 2018b, 2018c) conduct field experiments focused on testing ways to affect payments for these consumers. The majority of credit card users in the UK do not have automatic payments set up (Financial Conduct Authority, 2016), with the majority of credit card payments made manually (typically online) making this the most important payment channel (Adams et al., 2018c).
   ii.  Second we remove respondents who had a zero statement balance which removes another 241 respondents. We did this because the population of interest is those who have payments due against their cards and so are active credit card users.

Applying these restrictions leaves us with 3,044 respondents. Table 2 shows that these consumers typically revolve just over £2,000 in debt on their credit card. Just over one in ten only pays exactly the contractual minimum almost every month.

---

[3] The prize draw offered two £500 Amazon gift vouchers and fifteen £100 Amazon gift vouchers. Due to UK marketing research regulations entry into this prize draw was not conditional on completing the survey.

Our survey includes a UK Office for National Statistics (ONS) question used in its Wealth & Assets Survey asking respondents how well they are keeping up with bills and commitments.[4] This has been previously shown by Gathergood & Guttman-Kenney (2016) to be closely-related to measures of subjective wellbeing and highly correlated with other measures of financial distress.

We group responses into three categories. 47.4% of respondents report no financial distress, 40.7% have some financial distress - reporting keeping up with payments but regarding it as an occasional struggle and the remaining 11.9% are clearly distressed - reporting being in a constant struggle to keep up with payments, falling behind with payments or having real financial problems.

## Financially-distressed minimum payments

We examine the actual credit card payments of consumers who gave consent to match their survey responses and find approximately one in four of cards paid at, or very close to, the contractual minimum amount. Figure 1 shows the distribution of payments and that most payments fall at or close to the minimum (except for a group paying in full). This distribution is consistent with earlier research using large, administrative datasets of credit card payments in the UK (FCA, 2016) and the US (Keys & Wang, 2016).

**Figure 1: Distribution of actual credit card payments (as a % of statement balance)**

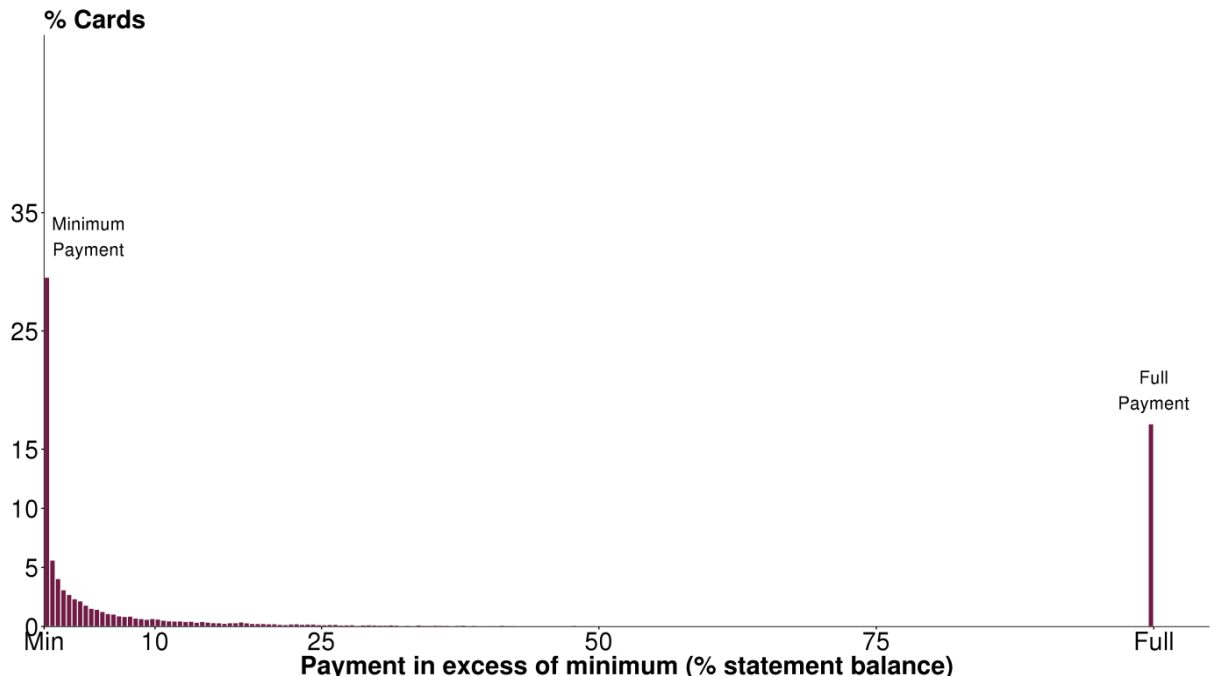

Notes: 0.5 percentage point buckets of payments. Only including observations where statement balance is positive and a payment was made against that statement (2,884 out of 3,044 observations).

---

[4] Which of the following statements best describes how well you are keeping up with your bills and credit commitments at the moment? Are you: Keeping up with all of them without any difficulties; Keeping up with all of them, but it is a struggle from time to time; Falling behind with some of them; Having real financial problems and have fallen behind with many of them; Don't have any commitments.

Our survey allows us to evaluate whether the clustering of payments at and close to the minimum is driven by economic forces. One hypothesis is that these consumers are liquidity constrained (Agarwal, Liu, & Souleles, 2007; Agarwal et al., 2015; Agarwal et al., 2017; Gross & Souleles, 2002; Leth-Petersen, 2010) to such an extent that they cannot afford to pay more than the minimum.[5] If this is the case, these consumers may want to pay more but be unable to.

As an initial test of this hypothesis, we examine how the distribution of respondents' actual credit card payments varies with self-reported financial distress. If economic forces are driving minimum payment behaviour we would expect consumers making minimum payments to be largely those reporting financial distress. We do not find this to be the case.

Figure 2 displays the distribution of payments for consumers reporting no financial distress (Panel A), with some financial distress (Panel B) and most distressed (Panel C). There is a high prevalence of minimum payments, irrespective of financial distress. The shapes of these distributions are also similar, with spikes of payments at the minimum and a mass of payments close to it. We find a natural ordering where higher proportions of consumers make minimum payments (and lower full payments) as we move across consumers with increasing financial distress. Yet in a given month more than one in four people who report no financial distress make a minimum payment.

These results provide evidence that the observed distribution of payments is not simply explained by economic forces. A behavioural phenomenon such as the anchoring effect clearly plays an important role. We do not suggest that liquidity constraints do not matter – they clearly do, as shown by the differing rates of minimum payments by financial distress - just that they do not appear to explain why so many consumers are paying at or close to the minimum.

We are therefore encouraged that it is possible for many consumers who currently pay the minimum to increase their credit card payments and reduce their credit card debt. With such a motivation in mind we proceed to testing an intervention designed to do just this – helping consumers not in financial distress pay down debt faster without forcing constrained individuals to make higher payments they may struggle to meet.

---

[5] Another hypothesis may be consumers are repeatedly only making minimum payments due to inertia of having automatic minimum payments set up which do not require the consumer to actively engage with their credit card (Sakaguchi, Stewart & Gathergood, 2018). Such behaviour is explored in Adams, Guttman-Kenney, Hayes, Hunt, Laibson & Stewart (2018b, 2018c) which respectively test the ability of disclosure and choice architecture to increase credit card payments for people who would otherwise be on automatic minimum payments.

**Figure 2: Distribution of actual credit card payments (as a % of statement balance), split by whether consumers report financial distress**

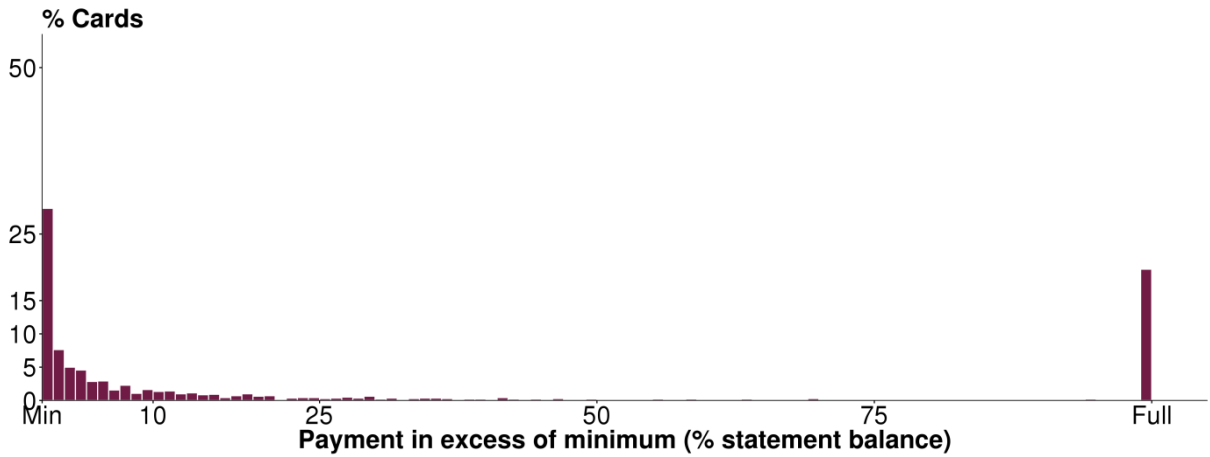

Panel A. No Financial Distress (N=1,400)

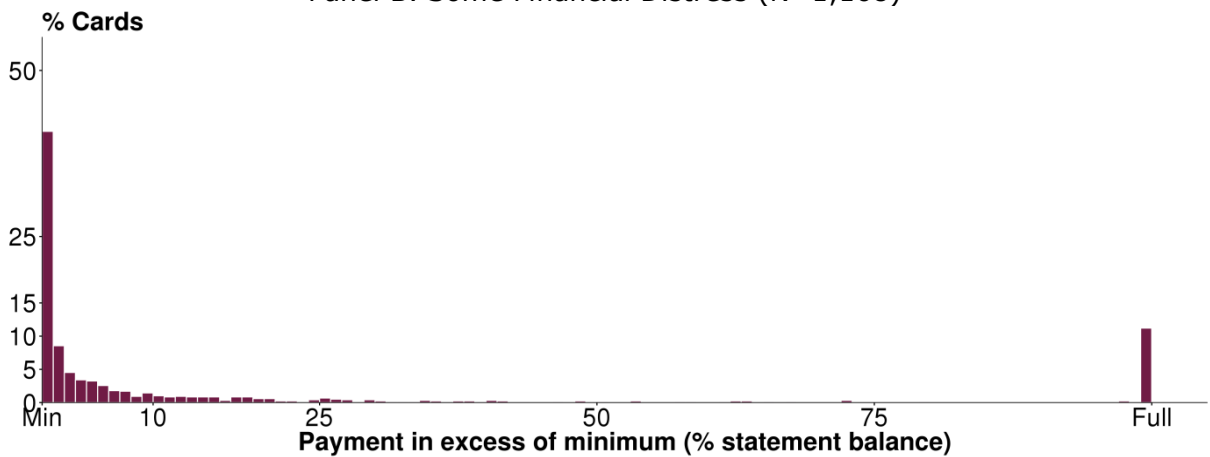

Panel B. Some Financial Distress (N=1,168)

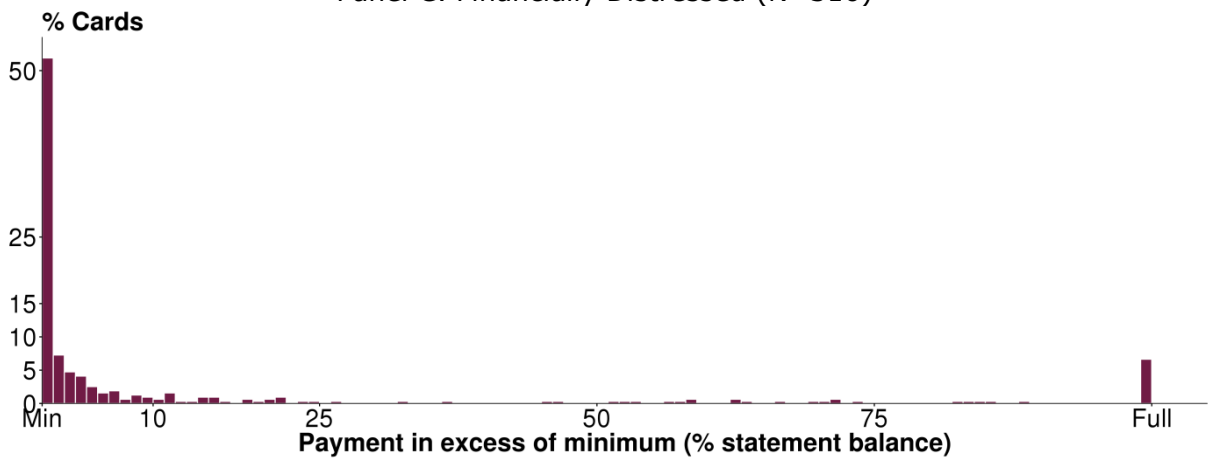

Panel C. Financially Distressed (N=316)

Notes: 1 percentage point buckets of payments. Only including observations where statement balance is positive and a payment was made against that statement (2,884 out of 3,044 observations).

# 3   De-anchoring experiment

## Experimental Design

We trial an intervention similar to that used in Stewart (2009) and subsequent papers (Jiang & Dunn, 2013; Navarro-Martinez et al., 2011). This intervention makes minimum payments less salient to enable consumers to make a de-anchored payment choice.

Our experiment presented consumers with a screen replicating that actually used for making manual credit card payments online. Consumers were presented with a hypothetical bill asking them to imagine this was their actual bill and, considering their actual financial situation, enter how much they would pay. Figure 3 shows this interface.

We use a randomised controlled trial (RCT) methodology where we randomly allocate consumers to each of the cells shown in Table 1, with equal probability. Half of consumers saw the minimum payment amount with a radio button option to pay only the minimum payment.[6] This control group replicates the actual current payment environment. The other half of consumers were assigned to the treatment group. They did not see the minimum payment amount or minimum payment radio button option. Comparing the treatment and control group enables us to measure the effect of minimum payments being prominently displayed as a payment option. Table 3 shows the randomisation was successful, with balance across observable behaviour between the two groups. This enables us to infer differences between treatment and control groups as causal estimates of the effect of the intervention.

In both scenarios, if a consumer input an amount less than the minimum payment, a prompt appeared which displayed the minimum payment amount. After receiving the prompt consumers were still able to pay an amount less than the minimum if they wanted to (eg if they could not afford the minimum payment but still wanted to pay something). UK credit card lenders already display this kind of payment prompt. The prompt is especially important for the treatment group, as otherwise consumers could easily choose to pay an amount less than the minimum unintentionally.

We also randomised the statement balance scenario shown to consumers. We used low and high balance scenarios taken from the 25th and 75th percentiles of the target samples actual, non-zero statement balances - taken from their seventh credit card statement after card opening. This enables us to better replicate different balances and also observe how the treatment effects of displaying the minimum payment vary based on the balance a consumer is shown.

---

[6] Consumers could also input the minimum amount if they typed this into the free text box payment option.

**Figure 3: Payment screen presented to control (Panel A) and treatment (Panel B) groups in hypothetical low balance scenario**

A. Control

B. Treatment

Notes: High balance scenario replaces statement balance with £3,217.36 and the minimum with £72.38.

**Table 1: Experimental Design**

| Data unit | Minimum Payment Displayed | No Minimum Payment Displayed |
|---|---|---|
| **Low Balance Scenario** (£532.60 statement, £11.98 minimum payment) | CONTROL$^L$ | TREATMENT$^L$ |
| **High Balance Scenario** (£3,217.36 statement, £72.38 minimum payment) | CONTROL$^H$ | TREATMENT$^H$ |

## Empirical Methodology

Our experiment has one observation for each survey response ($i$). The specification used to derive average treatment effects (ATEs) is displayed in **Equation 1**. We generally estimate one equation by pooling observations from the low and high balance scenarios. But for some analysis we split out the two scenarios to explore how effects vary by balance.

**Equation 1**

$$y_i = \alpha + \delta TREATMENT_i + \varepsilon_i$$

Each regression includes a constant ($\alpha$) and ($\delta$) is the coefficient for the average treatment effect of the intervention. $TREATMENT_i$ is a binary variable taking a value of 1 if the respondent is assigned to the treatment group.

The outcomes ($y_i$) we examine are:

1. full payment (binary)
2. minimum payment (binary)
3. below minimum payment (binary)
4. payment amount (% statement balance)

Measures 1-3 are binary outcomes where we estimate **Equation 1** using logistic regressions. Measure 4 is a continuous measure estimated using OLS regressions with robust standard errors. Payment amount is normalised as a percent of statement balance to enable cleaner aggregation between low and high balance scenarios.

In line with Benjamin et al. (2018) we regard a p-value of 0.005 as the threshold for statistical significance. But we also highlight where results are 'suggestively significant' at the 0.01 and 0.05 levels. This approach reduces the false positive rates by applying a tougher threshold for statistical significance than the traditional 0.05 level. Other methodologies such as applying Bonferroni adjustments or familywise error corrections (List, Shaikh, & Xu, 2016) could similarly limit the potential for p-hacking (Simmons, Nelson, & Simonsohn, 2011).

## Average treatment effects

Overall, we find our intervention reduces minimum payments, increases average and full payments and does not increase below minimum payments. These results are shown in Table 4.

We find that our intervention causes a noticeable increase in the likelihood of paying debt in full. This surprising effect was found in Stewart's (2009) initial experiment on credit card anchoring effects. Our intervention increases the likelihood of paying debt in full by 4.4-9.9 percentage points (95% confidence interval). This is an average treatment effect increase of almost fifty percent, compared to the control group where 14.8% select full payments. Such an increase in full payments could indicate that that, when de-anchored from the minimum, consumers use the full payment amount as a target (Bartels & Sussman, 2016).

In line with prior studies (Jiang & Dunn, 2013; Navarro-Martinez et al., 2011; Stewart, 2009) we also find that our intervention dramatically reduces the likelihood of consumers choosing to pay exactly the contractual minimum. The intervention effectively eradicates minimum payments: causing a 24.7-29.3 percentage point decline in minimum payments and leaving less than 2% paying exactly the minimum in the treatment group. This is an average treatment effect decrease of 95% compared to the control group, where 28.5% select minimum payments.

A potential concern with our intervention is that a de-anchored choice makes it possible for consumers to pay less than the minimum payment and accidently miss payments as a result. We examine the role of the minimum payment prompts (seen by consumers who initially choose to pay an amount less than the minimum) in preventing such accidental too-low payments. In the treatment group, 4.7% of consumers initially select an amount less than the contractual minimum. This is far higher than in the control group, where 0.9% consumers make this initial choice. But after the prompt just under 1% of consumers in the treatment group select to pay less than the minimum. This is not significantly different from the control group. So it appears a simple prompt (as is already currently used by UK credit card lenders) is effective at preventing a potential undesirable effect from de-anchoring.

The intervention both reduces minimum payments and increases full payments. It also results in consumers being more likely to pay particular payment amounts along the distribution rather than amounts very close to the contractual minimum. Consumers are especially likely to choose prominent round numbers but not simply select the lowest round number above the contractual minimum. This tendency to choose round-numbered payments is in line with prior findings of Sakaguchi, Stewart and Gathergood (2018). In the treatment group, 69.6% of consumers who do not choose to pay the minimum or full amount are choosing to pay £50, £100, £150, £200, £250, £300 or £500. £100 is the most common figure selected, with 34.6% of consumers choosing it.

When we examine the overall effect of the intervention on payments (as a percent of statement balance) we find large effects. The intervention causes 8.8-13.8 percentage point increase in the size of payments. This equates to an average increase of over 44% on the control group average payment.

The intervention does not just shift up payments up but alters the distribution of payments, making them less concentrated around the minimum. This is seen in Figure 4 which compares the distribution of payments for control and treatment groups.

**Figure 4: Effect of treatment on distribution of payments (as a % of statement balance) in hypothetical scenarios**

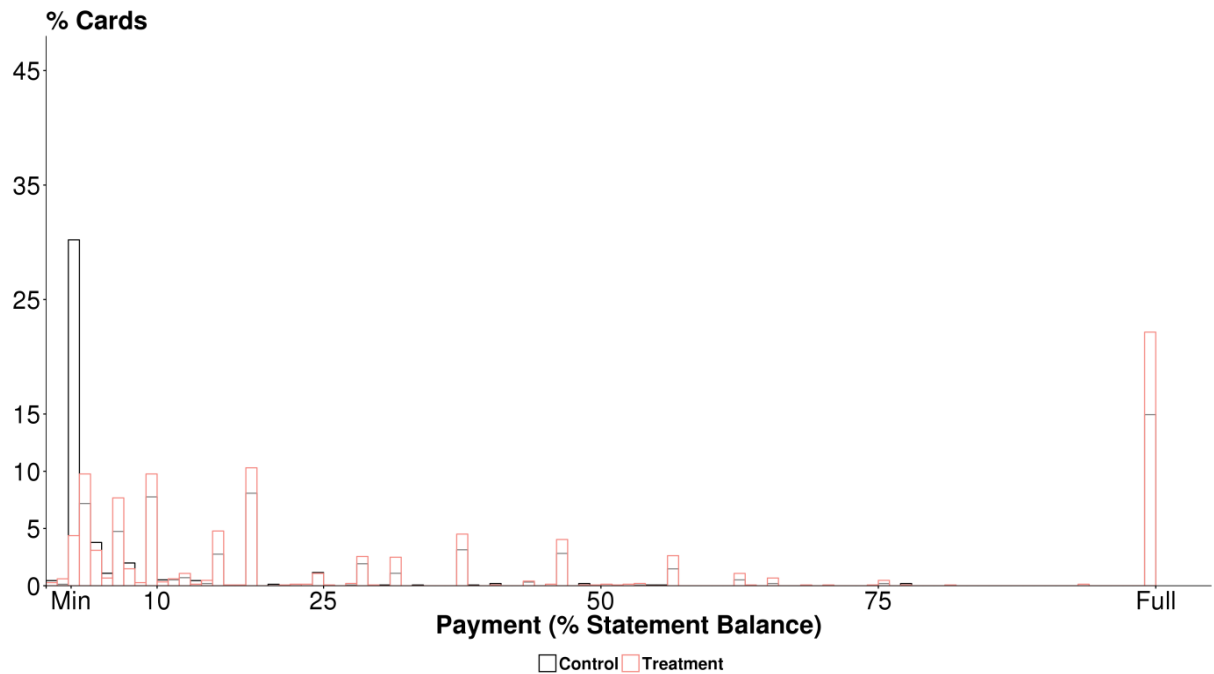

### Average treatment effects across balance scenarios

Table 5 looks at how our average effects of the intervention vary between our low balance scenario of £532 and our high balance scenarios of £3,217. This helps us to understand whether anchoring (and our intervention) affects consumers with larger credit card debts more or less than those with smaller ones.

We find the intervention has similar effects in increasing the likelihood that consumers pay in full in both low and high balance scenarios. Full payments increase by 2.7-11.1 percentage points in the low balance scenario and 3.8-10.8 percentage points in the high balance scenario. Because more consumers select full payments in the low than in the high balance scenario, the average percentage impact compared to their control groups vary between the two: 36% in the low and 69% in the high.

In both the low and high balance scenarios the intervention substantial reduces how often consumers only pay the minimum. The intervention reduces minimum payments by 14.6-20.0 percentage points for the low balance scenario and 33.1-40.3 percentage points for the high balance scenario. These are over 90% average decreases compared to the control group's choice of minimum payments. We find no evidence in either scenario of the intervention resulting in consumers being more likely to pay less than the minimum.

The effect of the intervention on average payments is nearly identical across the two scenarios (with a point estimate of 11%). In the low balance scenario, average payments are increased by 7.8-14.9 percentage points and in the high balance scenario by 7.7-14.4 percentage points. These are respectively 33% and 66% average increases in the control group's payments. In monetary terms, this is an average increase in payments of £60 in the low balance scenario and £355 in the high balance scenario.

So it appears that while there are some differences in how the intervention affects low and high balance scenarios the results are still broadly consistent.

# 4   Are hypothetical responses realistic?

We are reassured that our experiment finds effects that are consistent with previous studies. But our experiment, along with much of the prior literature, is based on hypothetical rather than actual payment decisions.[7] A common critique of such experiments is that the respondent has little reason to take a hypothetical task seriously. As a result, their responses may not exactly match their actions if they faced a similar choice in the real-world (Galizzi & Navarro-Martinez, 2018; Kessler & Vesterlund, 2015; Levitt & List, 2006, 2007) though that does not mean such experiments are not helpful in understanding how and why consumers make decisions outside the experiment (Camerer, 2015; Kessler & Vesterlund, 2015).

We measure the effects of the intervention in our experiment as the difference between treatment and control groups, which should cancel out systematic bias in hypothetical responses. However, to evaluate the scale of the treatment effects we often compare to the baseline levels of the control group. As the control group is intended to replicate actual choices in the real-world, we can examine the realism of the control group responses by comparing them with actual payment behaviour.

We do this using three methodologies: '*Visual*', '*Correlational*' and '*Predictive*'.

## Visual

The '*Visual*' approach compares the distribution of hypothetical payments in the experiment's control group to actual payments by customers with balances similar to those in the experiment. We use administrative data on the full target survey sample, which involves over 180,000 consumers, including respondents and non-respondents to the survey. We limit the comparison to consumers' payments in months when they made a payment against a statement, did not have automatic payments set up, and had similar statement balances to those in the hypothetical scenarios.

For the low balance scenario (£532.60) we chose the balance region of £500.00 to £549.99. For the high balance scenario (£3,217.36) it is £3,000.00 to £3,499.99. These regions were chosen because of the importance of the left-most digits in consumer decisions (Gabaix, 2017; Lacetera, Pope, & Sydnor, 2012; Thomas & Morwitz, 2005). We use a consumer's most recent actual statement, so there is at most one observation per consumer. This leaves 1,200 and 3,218 actual payment decisions from administrative data close to the low and high balance scenarios respectively. We compare these with the just over 1,050 responses in each of the hypothetical scenarios. We use the same restrictions as earlier, except that we do not exclude consumers who did not give consent to match survey responses to administrative data. This is because we are not using the match for this exercise. Figure 5 shows histograms of the hypothetical payments in the experiment compared to actual payments - separately for low (Panel A) and high balance (Panel B) scenarios. The low and high balance scenarios have different distributions of actual payments, such as more payments closer to the minimum in the high than the low balance scenario. We observe a similar pattern in hypothetical payment choices.

---

[7] The FCA attempted to trial this in the field with UK lender(s) on actual payments but no lender was willing and/or able to do so.

Responses to the high balance scenario show a similar shape to the distribution of actual payments. Compared to the hypothetical scenarios, a higher proportion of actual payments are minimum payments and fewer are full payments. We also observe larger spikes in the middle of the distribution for the hypothetical than actual payments. These spikes are round number payments. Round number payments are also present in the actual payments, but are smoothed out when payments are converted to percentage-of-balance. So, for example, £500 is a 16% repayment in the hypothetical scenario but 14-17% for actual payments as balances vary between £3,000.00 and £3,499.99. We see more noticeable spikes on the low balance scenario, where the distribution looks less close to actual payments. The proportion of consumers choosing full payments in the low balance scenario is very similar in hypothetical and actuals, as is the proportion choosing to pay the contractual minimum.

While our analysis is focused on respondents without automatic payments, another cross-check looks at the choices of respondents with automatic payments. This consumer segment is a useful cross-check as they have very clear payment patterns. Consumers with automatic full payments commonly pay debt in full, those with automatic minimum payments commonly pay only the minimum. In our hypothetical scenarios we find consumers with automatic full payments mostly report paying in full and those with automatic minimum payments commonly selecting the minimum. 72.1% of consumers with automatic full payments select to pay in full in our hypothetical scenario. 44.1% of consumers with automatic minimum payments select to pay the minimum in our hypothetical scenario.

## Correlational

The '*Correlational*' approach looks at the sub-group of consumers in the control group who gave consent to match survey responses to administrative data and had similar balances to the hypothetical balances they were given (as used for the 'Visual' approach). This restricts the sample to 779 and 774 responses respectively in the low and high scenarios. We can observe how these consumers' hypothetical choice compares to that same consumers' actual payment choice closest to that scenario (one observation per consumer). A crude correlational comparison of hypothetical payments in the experiment against their actual payments (both normalised as a percent of statement balance) reveals reasonable positive correlations of 0.366 and 0.419 for the low and high balance scenario respectively.

**Figure 5: Distribution of payments in control group compared to actual payments for hypothetical low balance (Panel A) and high balance (Panel B) scenarios**

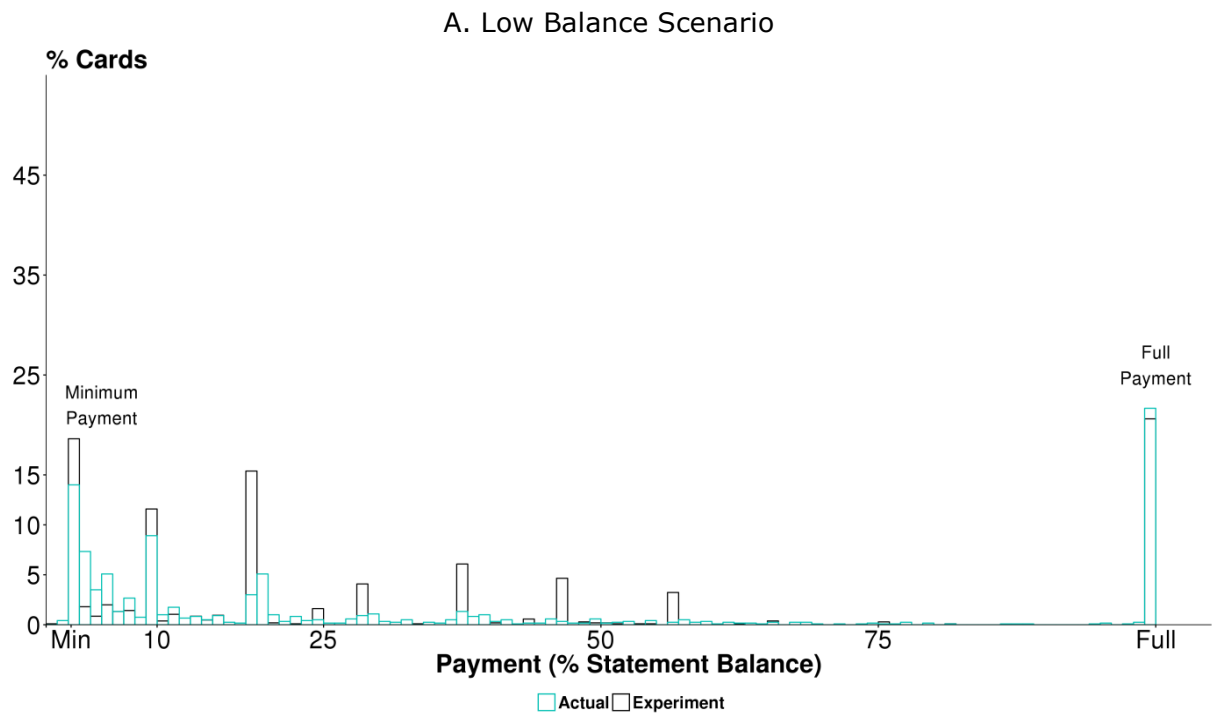

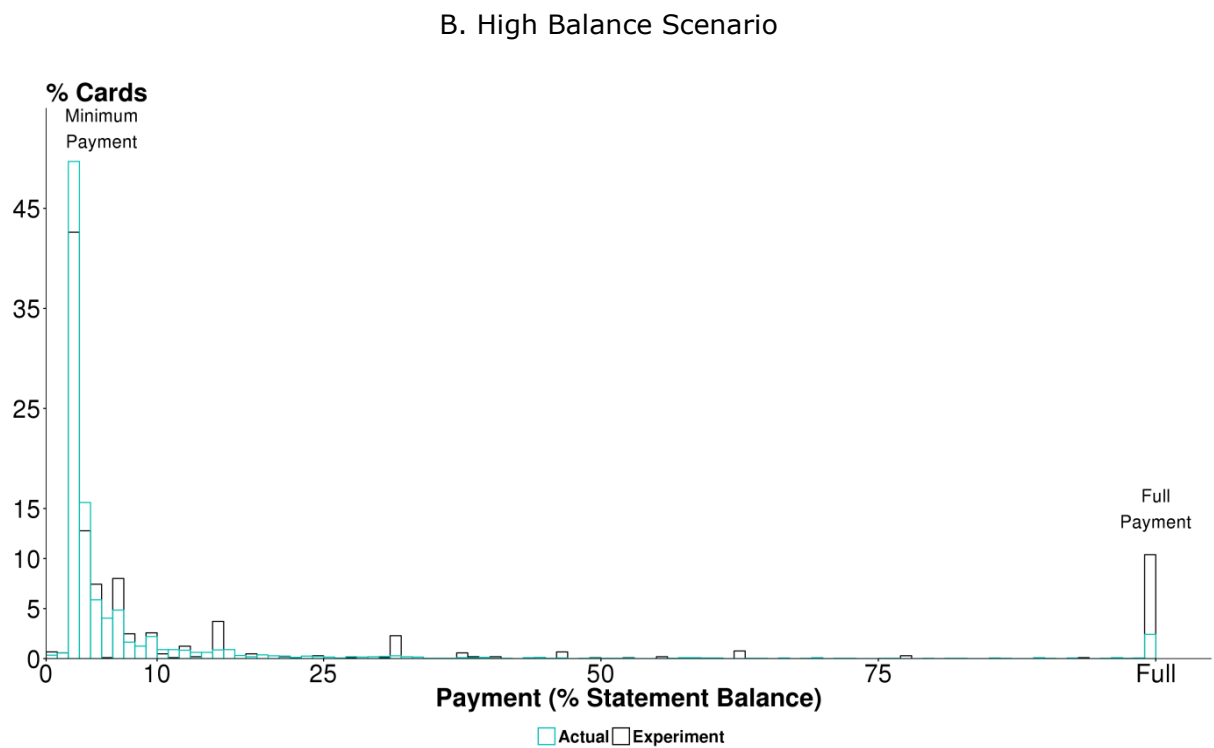

## Predictive

Finally, the '*Predictive*' approach uses the same group of consumers as the '*Correlational*' approach but uses a more sophisticated empirical methodology. We use an OLS regression model to examine how much of the variation in hypothetical payments can be predicted by consumers' actual payment behaviour. We treat actual payment as a percent of balance, log absolute difference in actual statement to hypothetical scenario, whether actual statement above hypothetical scenario, actual number of minimum payments and actual number of full payments. As shown in Table 6, this basic approach can predict (adjusted $R^2$) 15.8% and 14.2% of the variation in hypothetical payments in the low and high balance scenarios. Coefficients on consumers' actual payments are positive, large and highly statistically significant in both the low and high balance scenarios. A one percentage point increase in actual payments predicts a 0.24 percentage point increase in hypothetical payments in the low balance scenario. For the high balance scenario the prediction is slightly weaker at 0.16 percentage points. Finally, in both scenarios we observe that whether consumers have actually made many full payments in the past is highly predictive of their hypothetical payment choice.

By three different methodological approaches we have shown that hypothetical responses appear to be closely related to actual payment choices. This gives us reassurance that if our intervention was applied to the real world the impacts may be similar to those found in our hypothetical experiment.

# 5 Effects of de-anchoring across consumers

So far in our hypothetical experiment analysis we have focused on the average effects of the intervention. However, investigating the distribution of impacts could help us to better understand the types of consumers who would be most and least affected by the intervention if it were applied to actual credit card payment decisions. Given the large average treatment effects and size of our experiment, we have sufficient statistical power to analyse some potential basic heterogeneity in treatment effects. We examine how effects vary by self-reported financial distress and by a variety of measures of consumers' actual credit card payment behaviour. These are: the number of minimum payments and the number of full payments, and the average size of payments. We measure how effects vary by estimating **Equation 1** but adding an interaction term between the segment of interest and the treatment.

## Financial distress

We generally find the intervention most affects less financially-distressed participants, suggesting these consumers may have more capacity to increase payments than more distressed consumers (Table 7). The intervention results in a 5.4-14.7 percentage point (95% confidence interval) increase in full payments among consumers not in financial distress: an over 40% average increase on the control group. Effects on full payments for consumers with some financial distress are suggestively significant, and insignificant for financial distressed consumers. The effect on average payments for consumers with some and no financial distress are also substantial: 6.8-14.3 and 9.0-16.0 percentage point increases for consumers with some and no financial distress respectively. These are 58% and 35% average increases compared to the control group segments with some and no financial distress.

There is over a 90% reduction in consumers making minimum payments observed in each of the three financial distress segments (no distress, some distress or distressed). This is most dramatic for the financially-distressed consumer segment; 51.3-65.2% of the control group make minimum payments compared to just 2.0-8.8% in the treatment group. This financially-distressed group, however, does not have a statistically significant increase in payment size. This indicates that while they no longer choose the contractual minimum they would not make large increases in their payment.

The intervention does not cause any change in payments that are less than the minimum in any of the financial distress segments.

The effects on payments of the three groups can be seen in Figure 6. For the financially-distressed group (Panel C), the treatment causes more payments (which would have been expected to be at the minimum in the control group) to be just above the minimum and concentrated at 10% or less of the statement balance. For consumers with some financial distress (Panel B), there is a greater spread of payment values in the treatment group. For consumers not in financial distress (Panel A), much of the mass moves from one extreme – minimum payments – to the other – full payments. While these distributions appear to show a clear pattern, the group of consumers in financial distress is small. As a result, our confidence intervals are wide for this segment, meaning that we cannot conclusively say the treatment effects differ.

**Figure 6: Treatment effects on average payments (as a percent of statement balance), split by whether consumers report financial distress**

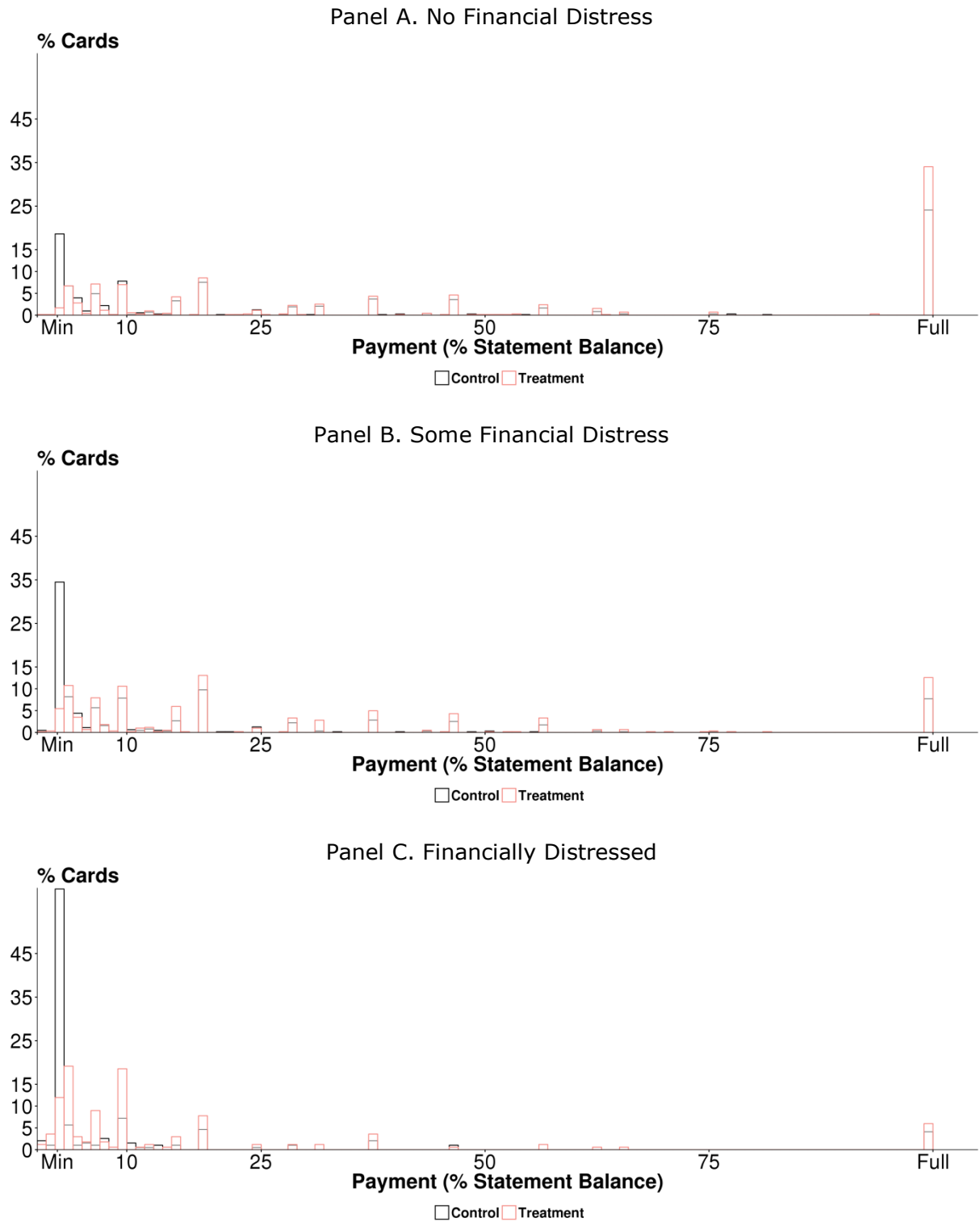

## Treatment effects by actual payment behaviour

We estimate the treatment effects on groups of consumers categorised by their actual payment behaviour.

### Full payments

We first look at full-payment behaviour, segmenting consumers by whether they made zero, a low (1-3) or high (4-7) number of full payments in the seven cycles of credit card payments we observed.[8] Regression results are shown in Table 8.

We find that the intervention increases the likelihood of consumers selecting full payments. This holds both for respondents who actually did not make full payments and those who made a low number of full payments. The likelihood of making a full payment increases by 3.0-10.1 percentage points among consumers who did not actually make full payments and 4.8-12.6 percentage points for those who only made a small number of full payments. These are average increases of 45% and 88% for the two groups, compared to the control group, where fewer than one in ten select full payments in our scenarios. We do not find a statistically significant effect of the intervention for consumers who actually made many full payments. This is perhaps unsurprising, given that approximately forty percent of this segment already select to pay in full in the hypothetical scenario.

Looking at the effects of the intervention on average payments shows broadly similar increases across the three segments of full payments. Average payments are increased by 8.1-15.5, 8.4-16.0 and 4.4-15.9 percentage points respectively for consumers with zero, a low and a high number of full payments.

### Minimum Payments

We next categorise consumers by whether they actually made zero, a low (1-3), or high (4-7) number of minimum payments in the seven cycles of credit card payments observed.[9] Table 9 shows estimated treatment effects for consumers categorised in this way. It shows that the intervention's effect of increasing full payments appears to be driven primarily by consumers who do not actually make minimum payments (ie segments of consumers who are making payments close to but not at the minimum in the control group make full payments in the treatment group). The intervention causes a 4.6-12.1 percentage point increase in full payments among consumers who made zero actual minimum payments (Table 9).

We observe a logical ordering in the responses of the control group when categorised by actual payment behaviour. Increasing proportions of consumers choose minimum payments if they have actually made no, some or many minimum payments. The intervention causes a 42.4-60.6 percentage point reduction in minimum payments among those who actually made many minimum payments (a 90% average change). Percentage point declines are smaller at 34.7-44.1 percentage points for consumers who make a low number of minimum payments and those who did not actually make minimum payments (15.0-20.0 percentage points). However, proportionally these changes are similarly, with relative declines in minimum payments of roughly 95% percent. It is not surprising that some consumers in the control group who did not make an actual minimum payments gave a hypothetical response that they would do so, as the hypothetical balance could be substantially higher than any of their actual balances.

---

[8] In the control group: 41.2% actually made zero minimum payments, 40.7% made a low number of minimum payments and 18.1% made a high number of minimum payments.

[9] In the control group: 61.4% made zero actual minimum payments, 29.7% made a low number of minimum payments and 8.9% made a high number of minimum payments.

We find statistically significant positive treatment effects on average payments for all three segments of consumers. The intervention's effects are largest for consumers who actually made a high number of minimum payments: a 9.1-25.6 percentage point or 142 percent average increase in payments. This compares to a 5.3-14.4 percentage point increase (52% average increase) for consumers with a low number of minimum payments and a 8.1-14.4 percentage point increase (36.8% average increase) for those who did not actually make minimum payments.

**Average Payments**

Finally, we segment consumers into three equally sized groups by their average actual payment, as a percent of balance, over seven cycles. We do this to understand where the consumers who the intervention appears to affect most are in the distribution of payments. For example, does it mostly shift consumers paying low amounts to high amounts or high amounts to even higher amounts?[10]

By doing this we can see that the intervention causes similar-sized reductions in minimum payments of over twenty percentage points across consumers with low, medium and high actual payments.

Estimated treatment effects on average payments are also similar in percentage points across these three segments. Average payments increase by 5.5-16.2, 5.9-16.7 and 9.5-20.2 percentage points for consumers actually making low, medium and high average payments. Compared to the control group's payment, this equate to average, relative increases in payments of 67.9, 52.4 and 54.1%.

---

[10] The thresholds used for allocating consumers to these three groups are thirds of the distribution of average payments across seven cycles (this average calculation excludes months where a consumer has a zero statement balance which would produce a zero denominator). These are average payments of 5.87% and 24.72% of average statement balances.

# 6   Concluding discussion

Our evidence demonstrates that how often consumers pay at or close to the minimum amount on their credit card does not appear to be fully explained by how financially distressed they are. Instead it appears that a behavioural phenomenon such as the anchoring effect is leading consumers to make low payments and so carry more credit card debt.

We test the effects of an intervention designed to address the anchoring effect by requiring consumers to make de-anchored choices of payment amounts. We build on the existing literature by using a larger and more representative sample, using multiple hypothetical balance scenarios, and linking actual payment behaviour.

Consistent with earlier literature, we find that in a hypothetical setting the intervention has large effects on the distribution of payments. It reduces minimum payments and increases both full payments and average payments. These results are consistent across low and high balance scenarios.

We use a variety of approaches to evaluate the external validity of our survey experiment by comparing hypothetical payments in the survey to administrative data on respondents' actual credit card payments. Across our tests, consumers appear to be taking the hypothetical task seriously, as hypothetical responses are consistent with consumers' actual payment choices. This gives us greater confidence in the external validity of the treatment effects found in our experiment. We conclude that, if applied to real-world payment decisions, results would likely go in the same direction. We would expect the relative distribution of impacts of across types of consumers as shown and the effects themselves may likely to be in, or close to, the confidence intervals we find, however, they may also change over time.

Our final contribution is to be able to examine how the effects of the intervention vary across consumers. While it appears that the intervention affects those not in financial distress the most, we do not have sufficient statistical power to say this conclusively. If someone has not got the money their payment choice is likely not primarily due to a behavioural bias but due to '*liquidity constraints*' – their budget constraint means they can only pay the minimum (or close to it). As liquidity constraints vary over time we would expect this intervention to result in consumers paying more in 'good times' when less constrained which results in less credit card debt and so lower contractual minimum payments required in 'bad times' when liquidity constrained. We can conclusively say that there appears to be strongly significant effects of the intervention increasing payments for those not reporting to be in financial distress.

We observe consumers who do not actually make minimum payments (eg selecting to pay somewhere in between the minimum and full amount) but do who actually make zero or few full payments, are increasingly likely to make full payments as a result of the intervention. By de-anchoring choices from the minimum in the new choice environment the full payment amount now seems to be acting as a target for consumers to aim towards.

# Annex 1: Tables

**Table 2: Summary statistics**

| Outcome | Mean |
|---:|:---:|
| Age (years) | 36.22 |
| Female (% cards) | 51.18 |
| Full Payments For 6+ Cycles (% cards) | 12.64 |
| Minimum Payments For 6+ Cycles (% cards) | 11.96 |
| Number of Full Payments Across Cycles 1-7 | 1.69 |
| Number of Minimum Payments Across Cycles 1-7 | 2.41 |
| Credit Limit (£) | 4174.80 |
| Credit Card Statement Balance (£) | 2250.38 |
| Credit Card Statement Balance Net of Payments (£) | 2038.91 |
| Keeping up, no problem (%) | 47.04 |
| Keeping up, occasional struggle (%) | 40.70 |
| Keeping up, constant struggle (%) | 7.79 |
| Falling behind with some (%) | 2.86 |
| Having real problems and fallen behind with many (%) | 1.22 |
| No commitments (%) | 0.39 |
| N | 3,044 |

**Table 3: Balance checks between control and treatment**

| Outcome | Mean (Control) | Mean (Treatment) | Mean Difference (Treatment-Control) | Percentage Difference Relative to Control | CI Lower (Treatment-Control) | CI Upper (Treatment-Control) | P Value | T Statistic | N (Control) | N (Treatment) |
|---|---|---|---|---|---|---|---|---|---|---|
| Age (years) | 34.35 | 34.68 | 0.33 | 0.97 | -0.51 | 1.17 | 0.435 | 0.781 | 1,559 | 1,485 |
| Female (% cards) | 53.18 | 51.85 | -1.32 | -2.49 | -4.87 | 2.23 | 0.465 | 0.731 | 1,559 | 1,485 |
| Full Payments For 6+ Cycles (% cards) | 11.42 | 10.71 | -0.71 | -6.22 | -2.94 | 1.52 | 0.532 | 0.625 | 1,559 | 1,485 |
| Minimum Payments For 6+ Cycles (% cards) | 3.21 | 3.23 | 0.03 | 0.78 | -1.23 | 1.28 | 0.969 | 0.039 | 1,559 | 1,485 |
| Number of Full Payments Across Cycles 1-7 | 1.65 | 1.56 | -0.09 | -5.37 | -0.24 | 0.07 | 0.263 | 1.119 | 1,559 | 1,485 |
| Number of Minimum Payments Across Cycles 1-7 | 0.95 | 0.96 | 0.02 | 1.64 | -0.10 | 0.13 | 0.788 | 0.268 | 1,559 | 1,485 |
| Credit Limit (£) | 3,248.59 | 3,488.18 | 239.59 | 7.38 | 32.04 | 447.14 | 0.024 | 2.263 | 1,559 | 1,485 |
| Credit Card Statement Balance (£) | 1,902.65 | 1,974.12 | 71.48 | 3.76 | -78.00 | 220.95 | 0.348 | 0.938 | 1,559 | 1,485 |
| Credit Card Statement Balance Net of Payments (£) | 1,701.86 | 1,769.72 | 67.86 | 3.99 | -78.25 | 213.97 | 0.362 | 0.911 | 1,559 | 1,485 |

*** P value < 0.005, ** < 0.01, * < 0.05.

**Table 4: Average treatment effects**

| | Outcome | Estimate | 95% Confidence Interval | Percentage Change | P Value | Degrees of Freedom | Adjusted R Squared |
|---|---|---|---|---|---|---|---|
| 1 | Full Payment | 0.0714*** (0.014) | [0.0439, 0.0988] | 48.18% | 0.0000 | 3042 | 0.0090 |
| 2 | Minimum Payment | -0.27*** (0.0119) | [-0.2932, -0.2467] | -94.60% | 0.0000 | 3042 | 0.1955 |
| 3 | Below Minimum Payment | 0.003 (0.0032) | [-0.0033, 0.0093] | 46.88% | 0.3495 | 3042 | 0.0032 |
| 4 | Payment (% Statement Balance) | 0.1131*** (0.0129) | [0.0878, 0.1384] | 44.34% | 0.0000 | 3042 | 0.0242 |

*** P value < 0.005, ** < 0.01, * < 0.05.

**Table 5: Average treatment effects separately for low and high balance scenarios**

| Outcome | Scenario | Estimate | 95% Confidence Interval | Percentage Change | P Value | Degrees of Freedom | Adjusted R Squared |
|---|---|---|---|---|---|---|---|
| Full Payment | 532.60 | 0.0688*** (0.0213) | [0.0271, 0.1105] | 36.06% | 0.0012 | 1534 | 0.0064 |
| | 3217.36 | 0.0727*** (0.0179) | [0.0375, 0.1078] | 68.98% | 0.0001 | 1506 | 0.0135 |
| Minimum Payment | 532.60 | -0.1727*** (0.0138) | [-0.1998, -0.1456] | -97.74% | 0.0000 | 1534 | 0.1852 |
| | 3217.36 | -0.3672*** (0.0185) | [-0.4035, -0.3309] | -93.06% | 0.0000 | 1506 | 0.2219 |
| Below Minimum Payment | 532.60 | 0.0026 (0.0019) | [-0.0010, 0.0063] | NA% | 0.1567 | 1534 | 0.0930 |
| | 3217.36 | 0.0036 (0.0062) | [-0.0086, 0.0157] | 27.91% | 0.5632 | 1506 | 0.0015 |
| Payment (£) | 532.60 | 60.30*** (9.69) | [41.30, 79.30] | 33.02% | 0.0000 | 1534 | 0.0240 |
| | 3217.36 | 355.00*** (55.31) | [246.60, 463.40] | 66.09% | 0.0000 | 1506 | 0.0260 |
| Payment (% Statement Balance) | 532.60 | 0.1132*** (0.0182) | [0.0775, 0.1489] | 33.01% | 0.0000 | 1534 | 0.0240 |
| | 3217.36 | 0.1103*** (0.0172) | [0.0766, 0.1440] | 66.05% | 0.0000 | 1506 | 0.0260 |

\*\*\* P value < 0.005, \*\* < 0.01, \* < 0.05.

**Table 6: Prediction of hypothetical payments using actual payment behaviour**

A. Low Balance Scenario (£532.60)

| Variable | Estimate | 95% Confidence Interval | P Value |
|---|---|---|---|
| Intercept | -0.0015 (0.0633) | [-0.1256, 0.1227] | 0.9817 |
| Actual Payment (% Statement Balance) | 0.2362*** (0.0535) | [0.1312, 0.3411] | 0.0000 |
| Log Difference in Actual-to-Hypothetical Balance | 0.0430*** (0.0092) | [0.0250, 0.0610] | 0.0000 |
| Actual Balance Above Hypothetical Balance | 0.0012 (0.0325) | [-0.0626, 0.065] | 0.9702 |
| Actual Number of Minimum Payments - low | -0.0473 (0.0272) | [-0.1005, 0.0059] | 0.0819 |
| Actual Number of Minimum Payments - high | -0.1119* (0.0450) | [-0.2002, -0.0237] | 0.0132 |
| Actual Number of Full Payments - low | -0.0002 (0.0263) | [-0.0517, 0.0514] | 0.9952 |
| Actual Number of Full Payments - high | 0.1973*** (0.0502) | [0.0988, 0.2957] | 0.0001 |

Degrees of Freedom: 751

Adjusted $R^2$=0.1580

*** P value < 0.005, ** < 0.01, * < 0.05.

B. High Balance Scenario (£3217.36)

| Variable | Estimate | 95% Confidence Interval | P Value |
|---|---|---|---|
| Intercept | 0.0940 | [-0.0810, 0.2690] | 0.2926 |
|  | (0.0893) |  |  |
| Actual Payment (% Statement Balance) | 0.1566*** | [0.0793, 0.2340] | 0.0001 |
|  | (0.0395) |  |  |
| Log Difference in Actual-to-Hypothetical Balance | -0.0006 | [-0.0237, 0.0225] | 0.9575 |
|  | (0.0118) |  |  |
| Actual Balance Above Hypothetical Balance | 0.0331 | [-0.0175, 0.0837] | 0.1996 |
|  | (0.0258) |  |  |
| Actual Number of Minimum Payments - low | -0.0169 | [-0.0621, 0.0282] | 0.4623 |
|  | (0.0230) |  |  |
| Actual Number of Minimum Payments - high | -0.0680 | [-0.1372, 0.0011] | 0.0543 |
|  | (0.0353) |  |  |
| Actual Number of Full Payments - low | 0.0096 | [-0.0359, 0.0551] | 0.6795 |
|  | (0.0232) |  |  |
| Actual Number of Full Payments - high | 0.1821*** | [0.1031, 0.2610] | 0.0000 |
|  | (0.0403) |  |  |

Degrees of Freedom: 735

Adjusted $R^2$=0.1419

*** P value < 0.005, ** < 0.01, * < 0.05.

**Table 7: Average treatment effects by self-reported financial distress**

| Outcome | Levels | Estimate | 95% Confidence Interval | Percentage Change | P Value | Degrees of Freedom | Adjusted R Squared |
|---|---|---|---|---|---|---|---|
| Full Payment | No Distress | 0.1006*** (0.0238) | [0.0541, 0.1472] | 41.97% | 0.0000 | 3038 | 0.0866 |
| | Some Distress | 0.0437** (0.017) | [0.0104, 0.0770] | 56.61% | 0.0100 | 3038 | 0.0866 |
| | Distressed | 0.0238 (0.0227) | [-0.0207, 0.0683] | 65.93% | 0.2949 | 3038 | 0.0866 |
| Minimum Payment | No Distress | -0.1712*** (0.0142) | [-0.1989, -0.1434] | -98.39% | 0.0000 | 3038 | 0.2527 |
| | Some Distress | -0.3030*** (0.0194) | [-0.3410, -0.2649] | -93.87% | 0.0000 | 3038 | 0.2527 |
| | Distressed | -0.5286*** (0.0395) | [-0.606, -0.4512] | -90.75% | 0.0000 | 3038 | 0.2527 |
| Below Minimum Payment | No Distress | 0.0028 (0.0028) | [-0.0026, 0.0083] | 200.00% | 0.3085 | 3038 | 0.1162 |
| | Some Distress | 0.0002 (0.0039) | [-0.0075, 0.0080] | 4.26% | 0.951 | 3038 | 0.1162 |
| | Distressed | 0.0170 (0.0207) | [-0.0236, 0.0575] | 55.02% | 0.4117 | 3038 | 0.1162 |
| Payment (% Statement Balance) | No Distress | 0.1250*** (0.0179) | [0.0900, 0.1601] | 34.78% | 0.0000 | 3038 | 0.1157 |
| | Some Distress | 0.1054*** (0.0193) | [0.0676, 0.1432] | 57.60% | 0.0000 | 3038 | 0.1157 |
| | Distressed | 0.0612 (0.0358) | [-0.0090, 0.1314] | 62.01% | 0.0875 | 3038 | 0.1157 |

*** P value < 0.005, ** < 0.01, * < 0.05.

**Table 8: Average treatment effects by number of actual full credit card payments**

| Outcome | Levels | Estimate | 95% Confidence Interval | Percentage Change | P Value | Degrees of Freedom | Adjusted R Squared |
|---|---|---|---|---|---|---|---|
| Full Payment | Zero | 0.0653*** (0.0180) | [0.0300, 0.1006] | 74.97% | 0.0003 | 3038 | 0.0906 |
|  | Low (1-3) | 0.0871*** (0.0200) | [0.0480, 0.1262] | 87.63% | 0.0000 | 3038 | 0.0906 |
|  | High (4-7) | 0.0711 (0.0429) | [-0.0129, 0.1551] | 17.90% | 0.0972 | 3038 | 0.0906 |
| Minimum Payment | Zero | -0.2893*** (0.0191) | [-0.3266, -0.2519] | -93.47% | 0.0000 | 3038 | 0.2105 |
|  | Low (1-3) | -0.3098*** (0.0192) | [-0.3474, -0.2722] | -95.82% | 0.0000 | 3038 | 0.2105 |
|  | High (4-7) | -0.1375*** (0.0217) | [-0.1800, -0.0949] | -94.57% | 0.0000 | 3038 | 0.2105 |
| Below Minimum Payment | Zero | 0.0016 (0.0056) | [-0.0094, 0.0125] | 17.20% | 0.7804 | 3038 | 0.0389 |
|  | Low (1-3) | 0.0103* (0.0047) | [0.0010, 0.0196] | 643.75% | 0.0296 | 3038 | 0.0389 |
|  | High (4-7) | -0.0106 (0.0061) | [-0.0226, 0.0013] | -100.00% | 0.0816 | 3038 | 0.0389 |
| Payment (% Statement Balance) | Zero | 0.1176*** (0.0189) | [0.0806, 0.1546] | 61.73% | 0.0000 | 3038 | 0.121 |
|  | Low (1-3) | 0.1223*** (0.0194) | [0.0843, 0.1602] | 58.29% | 0.0000 | 3038 | 0.121 |
|  | High (4-7) | 0.1019*** (0.0293) | [0.0445, 0.1594] | 20.21% | 0.0005 | 3038 | 0.121 |

\*\*\* P value < 0.005, \*\* < 0.01, \* < 0.05.

**Table 9: Average treatment effects by number of actual minimum credit card payments**

| Outcome | Levels | Estimate | 95% Confidence Interval | Percentage Change | P Value | Degrees of Freedom | Adjusted R Squared |
|---|---|---|---|---|---|---|---|
| Full Payment | Zero | 0.0834*** (0.0193) | [0.0456, 0.1212] | 45.40% | 0.0000 | 3038 | 0.0299 |
|  | Low (1-3) | 0.0393 (0.0216) | [-0.0030, 0.0815] | 39.54% | 0.0684 | 3038 | 0.0299 |
|  | High (4-7) | 0.1003** (0.0379) | [0.0260, 0.1745] | 153.83% | 0.0081 | 3038 | 0.0299 |
| Minimum Payment | Zero | -0.1750*** (0.0127) | [-0.2000, -0.1500] | -96.37% | 0.0000 | 3038 | 0.2552 |
|  | Low (1-3) | -0.3942*** (0.0239) | [-0.4410, -0.3474] | -95.06% | 0.0000 | 3038 | 0.2552 |
|  | High (4-7) | -0.5149*** (0.0465) | [-0.6061, -0.4237] | -89.94% | 0.0000 | 3038 | 0.2552 |
| Below Minimum Payment | Zero | 0.0015 (0.0039) | [-0.0061, 0.0090] | 23.81% | 0.7052 | 3038 | 0.0072 |
|  | Low (1-3) | 0.0072 (0.0067) | [-0.0059, 0.0202] | 110.77% | 0.2832 | 3038 | 0.0072 |
|  | High (4-7) | -0.0001 (0.0102) | [-0.0200, 0.0199] | -1.39% | 0.9959 | 3038 | 0.0072 |
| Payment (% Statement Balance) | Zero | 0.1124*** (0.0163) | [0.0805, 0.1442] | 36.82% | 0.0000 | 3038 | 0.0544 |
|  | Low (1-3) | 0.0985*** (0.0234) | [0.0527, 0.1443] | 51.52% | 0.0000 | 3038 | 0.0544 |
|  | High (4-7) | 0.1732*** (0.0422) | [0.0906, 0.2559] | 142.67% | 0.0000 | 3038 | 0.0544 |

*** P value < 0.005, ** < 0.01, * < 0.05.

**Table 10: Average treatment effects by actual average credit card payments**

| Outcome | Levels | Estimate | 95% Confidence Interval | Percentage Change | P Value | Degrees of Freedom | Adjusted R Squared |
|---|---|---|---|---|---|---|---|
| Full Payment | Low | 0.0771*** (0.0200) | [0.0380, 0.1162] | 100.00% | 0.0001 | 3030 | 0.0592 |
|  | Medium | 0.0852*** (0.0289) | [0.0286, 0.1418] | 32.86% | 0.0032 | 3030 | 0.0592 |
|  | High | 0.0664*** (0.0214) | [0.0245, 0.1084] | 65.35% | 0.0019 | 3030 | 0.0592 |
| Minimum Payment | Low | -0.3538*** (0.0226) | [-0.3980, -0.3095] | -93.72% | 0.0000 | 3030 | 0.2171 |
|  | Medium | -0.1929*** (0.018) | [-0.2281, -0.1576] | -94.84% | 0.0000 | 3030 | 0.2171 |
|  | High | -0.2712*** (0.0204) | [-0.3112, -0.2313] | -96.44% | 0.0000 | 3030 | 0.2171 |
| Below Minimum Payment | Low | 0.0059 (0.0052) | [-0.0043, 0.0161] | 147.50% | 0.2549 | 3030 | 0.0183 |
|  | Medium | -0.0014 (0.0044) | [-0.0100, 0.0072] | -25.00% | 0.7500 | 3030 | 0.0183 |
|  | High | 0.0042 (0.0068) | [-0.0091, 0.0176] | 42.86% | 0.5346 | 3030 | 0.0183 |
| Payment (% Statement Balance) | Low | 0.12200*** (0.0217) | [0.0795, 0.1645] | 73.10% | 0.0000 | 3030 | 0.0876 |
|  | Medium | 0.1219*** (0.0216) | [0.0795, 0.1643] | 32.37% | 0.0000 | 3030 | 0.0876 |
|  | High | 0.1098*** (0.0217) | [0.0673, 0.1523] | 51.33% | 0.000 | 3030 | 0.0876 |

*** P value < 0.005, ** < 0.01, * < 0.05.

# Annex 2: References